\documentclass[a4paper,11pt]{article}
\pdfoutput=1 

\usepackage{jinstpub}

\title{Studies on tetrafluoropropene-based gas mixtures with low environmental impact for Resistive Plate Chambers}
\author[a,1]{A. Bianchi,\note{Corresponding author.}}
\author[a]{S. Delsanto,}
\author[b]{P. Dupieux,}
\author[a]{A. Ferretti,}
\author[a]{M. Gagliardi,}
\author[b]{B. Joly,}
\author[b]{S. P. Manen,}
\author[c]{M. Marchisone,}
\author[a]{L. Micheletti,}
\author[a]{A. Rosano}
\author[a]{and E. Vercellin}

\affiliation[a]{Universit\`a degli Studi di Torino and INFN, Sezione di Torino, Via Pietro Giuria 1, 10125, Torino, Italy}
\affiliation[b]{Clermont Universit\'e, Universit\'e Blaise Pascal, CNRS/IN2P3, Laboratoire de Physique Corpusculaire, BP 10448, F-63000 Clermont-Ferrand, France}
\affiliation[c]{Institut de Physique Nucl\'eaire de Lyon, Universit\'e Claude Bernard, 4 rue Enrico Fermi, 69622, Villeurbanne, France}

\emailAdd{antonio.bianchi@unito.it}

\abstract{
Gaseous detectors are widely used in high-energy physics experiments, and in particular at the CERN Large Hadron Collider (LHC), to provide tracking and triggering over large volumes. It has been recently estimated that Resistive Plate Chambers (RPC), used for muon detection, have the highest contribution on the overall greenhouse gas (GHG) emissions at the LHC experiments.

Gas mixtures for RPCs are mainly made of C$_{2}$H$_{2}$F$_{4}$, which is a greenhouse gas with a high environmental impact in the atmosphere. C$_{2}$H$_{2}$F$_{4}$ is already phasing out of production, due to recent European Union (EU) regulations, and its cost is expected to increase in the near future. Therefore, finding alternative gas mixtures made of gas components with a low Global Warming Potential (GWP) has become extremely important for limiting the GHG emissions as well as for economic reasons.

The novel hydrofluoroolefins are likely appropriate candidates to replace C$_{2}$H$_{2}$F$_{4}$ due to their similar chemical structures. This study is focused on the characterization of innovative gas mixtures with tetrafluoropropene HFO1234ze(E) (C$_{3}$H$_{2}$F$_{4}$) that is one of the most eco-friendly hydrofluoroolefins, thanks to its very low GWP. HFO1234ze(E)-based gas mixtures with the addition of Ar, N$_{2}$, O$_{2}$ and CO$_{2}$ are extensively discussed in this paper as well as the role of \textit{i}-C$_{4}$H$_{10}$ and SF$_{6}$ in such mixtures.

}


\proceeding{15$^{\text{th}}$ Topical Seminar on Innovative Particle and Radiation Detectors\\
  when 14-17 October 2019\\
  where Siena, Italy}

\begin{document}
\maketitle
\flushbottom

\section{Introduction}
Gaseous detectors are widely used in high-energy physics experiments, and in particular at the CERN Large Hadron Collider (LHC), to provide tracking and triggering over large volumes. It has been recently estimated that Resistive Plate Chambers (RPC), used for muon detection, have the highest contribution on the overall greenhouse gas (GHG) emissions at the LHC experiments \cite{capeans2017strategies_ok}. RPC detectors are operated with a large fraction (between 89\% and 95\%) of tetrafluoroethane C$_{2}$H$_{2}$F$_{4}$, whose trade name is R-134a. In addition, different concentrations of \textit{i}-C$_{4}$H$_{10}$ and SF$_{6}$ are used in RPC gas mixtures with the aim to optimize the detector performance. In the case of RPCs at the ATLAS and CMS experiments, the concentrations of \textit{i}-C$_{4}$H$_{10}$ and SF$_{6}$ are 4.5\% and 0.3\%, respectively, while the ALICE gas mixture contains 10.0\% \textit{i}-C$_{4}$H$_{10}$ and 0.3\% SF$_{6}$.

A greenhouse gas is classified according to its Global Warming Potential (GWP). This potential indicates the heat trapped by a gas in the atmosphere thus contributing to the greenhouse effect.  The GWP of CO$_{2}$ is equal to 1 by definition. On the contrary, the GWPs of C$_{2}$H$_{2}$F$_{4}$ and SF$_{6}$ are 1430 and 22800, respectively, whereas the GWP of \textit{i}-C$_{4}$H$_{10}$ is 3 \cite{europeanparliament_ok}. This implies that the GWP of the ATLAS/CMS gas mixture is \textasciitilde{}1430, while the GWP of the gas mixture used for the ALICE muon RPCs is \textasciitilde{}1351.

Recent regulations from the European Union (EU) impose a gradual limitation of the production and usage of fluorinated greenhouse gases (such as C$_{2}$H$_{2}$F$_{4}$) \cite{europeanparliament_ok}. This could lead to a progressive price increase of these gases. Therefore, finding gas mixtures made of low GWP components has become extremely important for limiting the GHG emissions in the atmosphere as well as for economic reasons. The new gas mixtures for RPCs must at least provide similar detector performance as the current ones.

In this paper we report the results of R\&D studies on eco-friendly gas mixtures for RPCs, used for the muon detection, at the LHC experiments.
\\
The paper is organized as follows. In section~\ref{sec:ecofriendly_gases} we discuss the approach to find alternative, environment-friendly, gas mixtures. The characterization of C$_{3}$H$_{2}$F$_{4}$-based gas mixtures with addition of the most common atmospheric gases is presented in section~\ref{sec:characterization_gases}, whereas the measurements of detector performance with gas mixtures of C$_{3}$H$_{2}$F$_{4}$ and CO$_{2}$ are reported in section~\ref{sec:characterization_anidridecarbonica}. Finally, conclusions are drawn in section~\ref{sec:conclusions}.

\section{RPC operation with environment-friendly gases} \label{sec:ecofriendly_gases}

RPC gas mixtures, currently used at the LHC experiments, are not environment-friendly because of the presence of greenhouse gases. More than 95\% of the total GWP of RPC gas mixtures at LHC is due to the presence of C$_{2}$H$_{2}$F$_{4}$.

The replacement of C$_{2}$H$_{2}$F$_{4}$ with tetrafluoropropene C$_{3}$H$_{2}$F$_{4}$ turns out to be a promising approach to find a more eco-friendly gas mixture \cite{Abbrescia_2016_He_ok, guida2016characterization_ok, liberti2016further_ok, bianchi2019r_ok}. Isomers of tetrafluoropropene are HFO1234ze(Z), HFO1234ze(E) and HFO1234yf. The most interesting isomer for the purposes of these R\&D studies is the trans-1,3,3,3-tetrafluoroprop-1-ene HFO1234ze(E) because it has a low boiling point (-19$\,$°C) and is not flammable below 30$\,$°C. The GWP of HFO1234ze(E) is 7 according to EU regulations \cite{europeanparliament_ok}, which is \textasciitilde{}200 times lower than that of C$_{2}$H$_{2}$F$_{4}$. More recent studies have revised the GWP of HFO1234ze(E) to less than 1 \cite{IPCCclimate_ok}. In addition, it has been estimated that the decay products of HFO1234ze(E) in the atmosphere have a negligible environmental impact \cite{javadi2008atmospheric_ok}.

R\&D studies on RPCs have been performed with HFO1234ze(E)-based gas mixtures. Hereafter, HFO1234ze(E) will be simply named as C$_{3}$H$_{2}$F$_{4}$ in order to be consistent with the chemical notation recognized by IUPAC\footnote{International Union of Pure and Applied Chemistry} and used for all other gases in this work. Detector performance of single-gap RPCs of the ALICE muon spectrometer \cite{alice1999alice_ok, arnaldi2000alice_ok} will be used as reference in this work. 

\subsection{Experimental set-up}
A detailed description of the experimental set-up is presented in a previous work \cite{bianchi2019characterization_ok}, while in this paper we only introduce its fundamental aspects.

R\&D studies on various gas mixtures are carried out through one small-size (50 $\times$ 50 cm\textsuperscript{2}, 2 mm gas gap) RPC with similar features as the ALICE muon RPCs \cite{alice1999alice_ok, arnaldi2000alice_ok}. The ALICE muon RPCs have electrodes in phenolic bakelite with a resistivity of $10^{9} \div 10^{10}$ $\Omega$ cm \cite{c_ok}, separated by a gas gap of 2 mm. The standard ALICE mixture is $89.7\%$ C\textsubscript{2}H\textsubscript{2}F\textsubscript{4}, $10.0\%$ \textit{i-}C\textsubscript{4}H\textsubscript{10} and $0.3\%$ SF\textsubscript{6} \cite{maxi_avalanche_ok}. The relative humidity of the gas mixture is kept constant at \textasciitilde{}37\% with the aim to prevent variations of the electrode resistivity \cite{water_ok, arnaldi2003aging_ok}.

In the experimental set-up, the detector is exposed to the cosmic-ray flux and equipped with the amplified front-end electronics FEERIC \cite{e_ok}. FEERIC amplifies signals up to $\pm$1 pC in the linear range \cite{e_ok}, which is the typical input charge expected for the RPCs in avalanche mode. The four adjacent strips covering the trigger area are read out on both ends: on one end the signals are discriminated by FEERIC with the aim to assess the efficiency, while on the other end the analogic signals are summed by a fan-in/fan-out module and then digitized by an oscilloscope (LeCroy WaveSurfer 510) in order to measure the signal amplitude \cite{bianchi2019characterization_ok}.

\begin{figure}[h]
    \centering
    \includegraphics[width=0.70\textwidth]{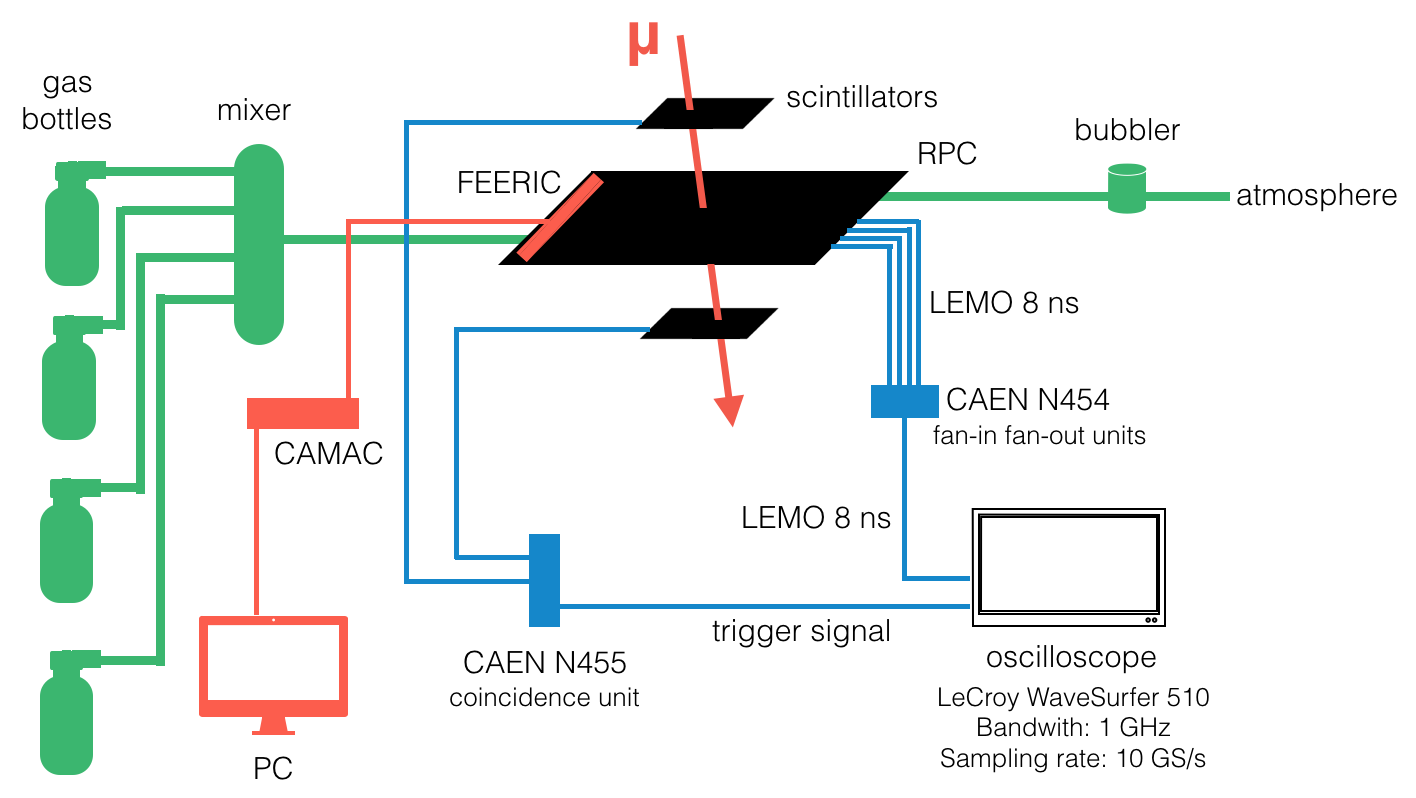}
    \caption{Sketch of the experimental set-up to measure the detector performance with C$_{3}$H$_{2}$F$_{4}$-based gas mixtures.}
    \label{fig:plot_setup}
\end{figure}

Figure \ref{fig:plot_setup} shows a sketch of the experimental set-up.

\subsection{Methodology}\label{sec:metodologia}

The new gas mixture with low GWP has to provide similar detector performance as the current one with the standard ALICE mixture. In this work, the working point, the fraction of streamers, the average charge of signals and the current stability are used to assess the performance of the tested gas mixtures.

The efficiency of the RPC detector is evaluated by analyzing the FEERIC response in each strip in the presence of a trigger signal \cite{bianchi2019characterization_ok}. The FEERIC discrimination threshold is \textasciitilde 130 fC (70 mV after the FEERIC amplification \cite{f_ok}). This threshold has been selected equal to the value used in one RPC of the ALICE muon identification system, equipped with FEERIC prototypes, with  fully satisfactory performance during the data taking a period from 2015 to 2018 \cite{f_ok}.

Signals with an amplitude higher than 18 mV are tagged as streamers \cite{bianchi2019characterization_ok}. Streamers are not desirable when RPCs are operated in avalanche mode because they reduce the rate capability of the detector \cite{sauli_ok} and may enhance ageing effects \cite{arnaldi2003aging_ok}.

\begin{figure}[h]
    \centering
    \includegraphics[width=0.80\textwidth]{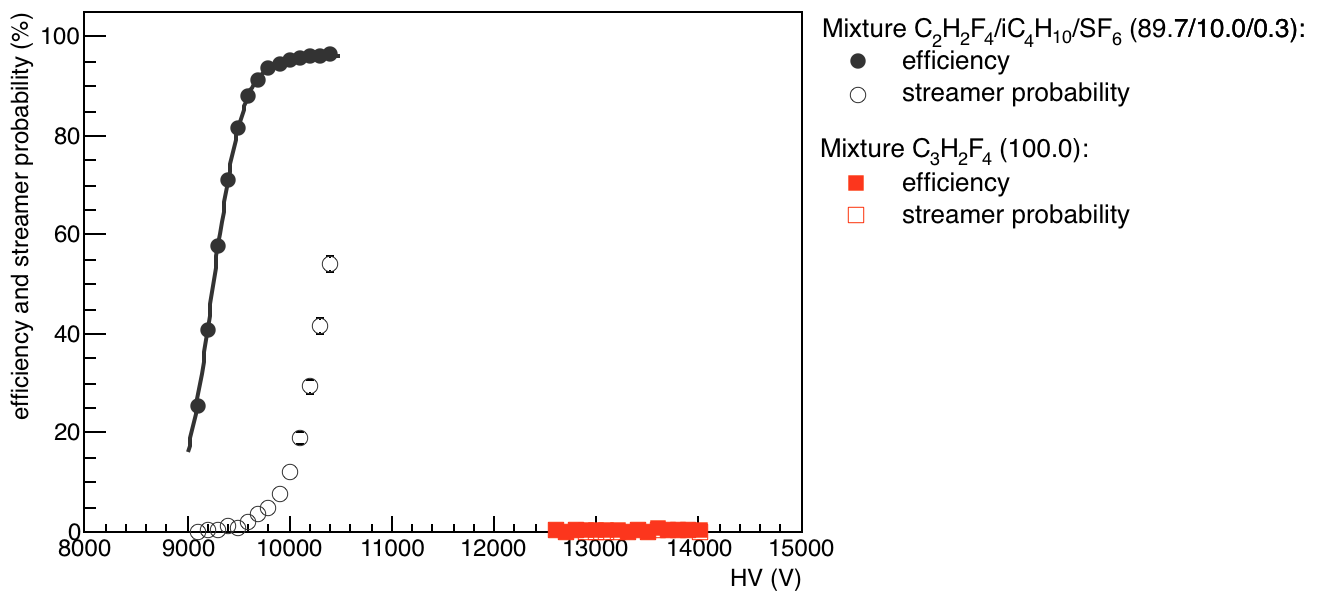}
    \caption{Efficiency curve and streamer probability of a RPC detector with the standard ALICE mixture and with pure C$_{3}$H$_{2}$F$_{4}$. Statistical error bars are hidden by markers.}
    \label{fig:plot_puro}
\end{figure}

In figure \ref{fig:plot_puro} the efficiency curve\footnote{The efficiency curves as a function of \textit{HV} are fitted by the following sigmoid function:
\begin{equation}
\epsilon(HV) = \frac{\epsilon\textsubscript{\normalfont{max}}}{1 + \exp[- \alpha \cdot (HV - HV_{0.5})]}
\end{equation}
where $\epsilon\textsubscript{\normalfont{max}}$ is the efficiency for $HV \xrightarrow{} \infty$, $HV_{0.5}$ is the $HV$ value for which the efficiency is equal to a half of $\epsilon\textsubscript{\normalfont{max}}$ and $\alpha$ is the slope at the inflection point \cite{bianchi2019characterization_ok}.} and the streamer probability versus \textit{HV} is shown for the ALICE mixture and for pure C$_{3}$H$_{2}$F$_{4}$. 

For the tested RPC, the efficiency plateau is reached at \textasciitilde{}9.8 kV when the detector is equipped with FEERIC and flushed with the ALICE mixture. The streamer probability is about 5\% at \textasciitilde{}9.8 kV. On the contrary, the direct replacement of the ALICE mixture with 100\% C$_{3}$H$_{2}$F$_{4}$ is not advisable because it leads to a too high operating voltage. Indeed, the RPC with 100\% C$_{3}$H$_{2}$F$_{4}$ is completely inefficient up to 14 kV, as shown in figure \ref{fig:plot_puro}.

In this work, C$_{3}$H$_{2}$F$_{4}$ is diluted with Ar, N$_{2}$, O$_{2}$ and CO$_{2}$ in order to reduce the operating voltage of the detector.

\section{Characterization of C$_{3}$H$_{2}$F$_{4}$-based gas mixtures with addition of Ar, N$_{2}$ or O$_{2}$} \label{sec:characterization_gases}

The most common atmospheric gases have been tested in C$_{3}$H$_{2}$F$_{4}$-based gas mixtures in order to lower the working point of RPCs. In particular, the study is here focused on Ar, N$_{2}$ and O$_{2}$ as their GWP is 0.

\begin{figure}[h]
    \centering
    \includegraphics[width=0.80\textwidth]{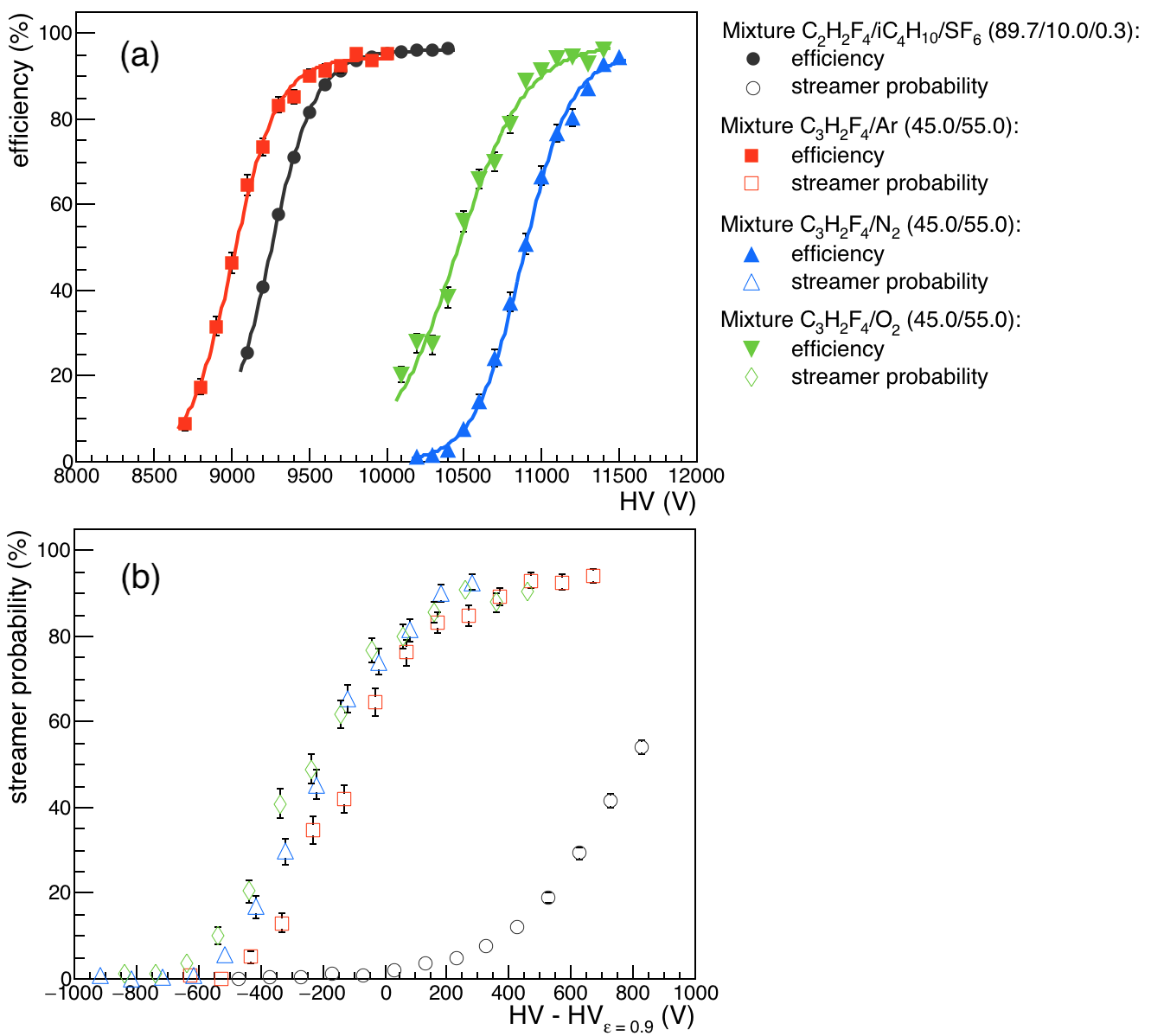}
    \caption{Efficiency (a) and streamer probability (b) of the RPC with 45\% C$_{3}$H$_{2}$F$_{4}$ and the remaining fraction of Ar, N$_{2}$ or O$_{2}$. In addition, the efficiency and the streamer probability of the RPC with the standard ALICE mixture are shown. Some statistical error bars are hidden by markers.}
    \label{fig:plot_misto}
\end{figure}

Figure \ref{fig:plot_misto} shows the experimental results of the efficiency and streamer probability curves of the RPC with 45\% C$_{3}$H$_{2}$F$_{4}$ and the remaining fraction of Ar, N$_{2}$ or O$_{2}$. As shown in figure \ref{fig:plot_misto}a, the RPC operating voltages with C$_{3}$H$_{2}$F$_{4}$-based gas mixtures and addition of Ar, N$_{2}$ and O$_{2}$ turn out to be extremely lower than that measured with pure C$_{3}$H$_{2}$F$_{4}$. In particular, the efficiency curve with Ar is at lower $HV$ values by \textasciitilde{}200 V in comparison to that with the standard ALICE mixture. On the contrary, the RPC working point with 55\% O$_{2}$ and 45\% C$_{3}$H$_{2}$F$_{4}$ is \textasciitilde{}1.0 kV higher than that with the standard ALICE mixture, while the use of N$_{2}$ in place of O$_{2}$ shifts the working point by 1.5 kV. In figure \ref{fig:plot_misto}b, the streamer probability curves are shown as a function of $HV$\textminus$HV_{\epsilon=0.9}$. $HV_{\epsilon=0.9}$ is the voltage where the efficiency is 90\% of its maximum value ($\epsilon_{max}$), and is obtained, for each mixture, by fitting the efficiency curve with a sigmoid function. No significant changes in the streamer suppression are observed if 55\% Ar, N$_{2}$ or O$_{2}$ is added to C$_{3}$H$_{2}$F$_{4}$. In fact, the streamer probability is always greater than \textasciitilde{}80\% at $HV$\textminus$HV_{\epsilon=0.9}$ of \textasciitilde{}200 V, while it is only \textasciitilde{}5\% in the standard ALICE mixture.

The average charge of RPC signals, induced in the read-out strips, has been measured at the working point in C$_{3}$H$_{2}$F$_{4}$-based gas mixtures with 55\% Ar, N$_{2}$ or O$_{2}$. In order to measure this quantity for each gas mixture under test, RPC signals are acquired by the oscilloscope and then integrated in a time window of 100 ns by including each signal totally. At the working point, the average charge is \textasciitilde{}2.6 pC with the standard ALICE mixture, while it turns out to be more than one order of magnitude higher in all other gas mixtures. In particular, the average charge is \textasciitilde{}66 pC and \textasciitilde{}68 pC in C$_{3}$H$_{2}$F$_{4}$-based gas mixtures with 55\% Ar and 55\% N$_{2}$, respectively. On the contrary, a lower average charge of \textasciitilde{}36 pC is measured with 55\% O$_{2}$. However, a current instability has been observed if the RPC operates with 45\% C$_{3}$H$_{2}$F$_{4}$ and 55\% O$_{2}$. In fact, the current at the working point increases by \textasciitilde{}60\% in 5 hours. The instability of the detector current has been observed in all C$_{3}$H$_{2}$F$_{4}$-based gas mixtures with O$_{2}$ in different concentrations, tested so far.

\section{Characterization of C$_{3}$H$_{2}$F$_{4}$-based gas mixtures with CO$_{2}$}\label{sec:characterization_anidridecarbonica}
Since Ar, N$_{2}$ and O$_{2}$ do not turn out to be suitable in C$_{3}$H$_{2}$F$_{4}$-based gas mixtures for RPCs, the use of CO$_{2}$ is here evaluated.

\begin{figure}[h]
    \centering
    \includegraphics[width=0.75\textwidth]{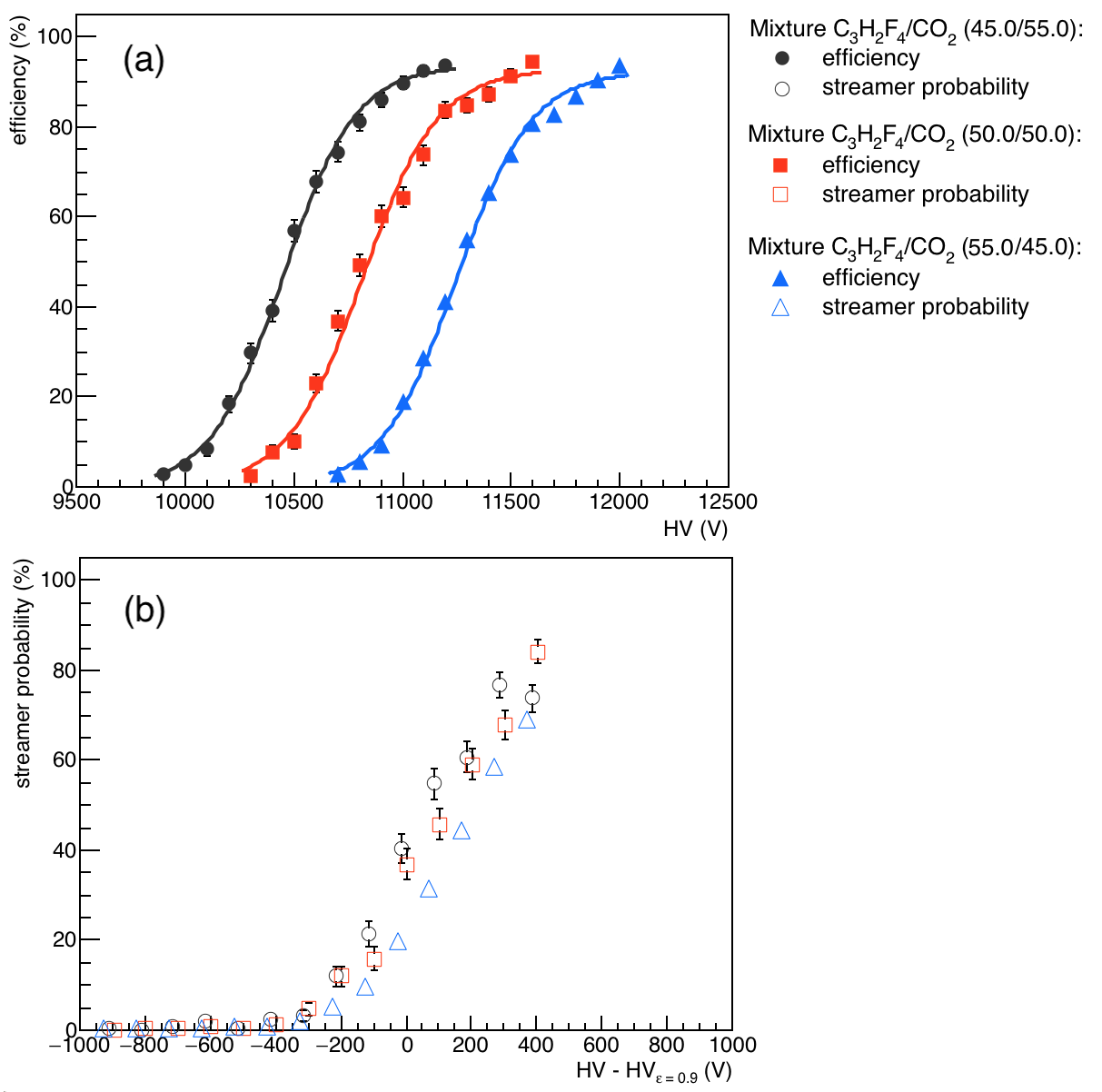}
    \caption{Efficiency (a) and streamer probability (b) of the RPC with C$_{3}$H$_{2}$F$_{4}$-based gas mixtures and 45\%, 50\% or 55\% CO$_{2}$. Some statistical error bars are hidden by markers.}
    \label{fig:plot_anidridecarbonica}
\end{figure}

Figure \ref{fig:plot_anidridecarbonica}a shows the efficiency curves of RPC at different ratios between C$_{3}$H$_{2}$F$_{4}$ and CO$_{2}$. 

If the gas mixture with 45\% C$_{3}$H$_{2}$F$_{4}$ and 55\% CO$_{2}$ is taken into consideration, the working point of the detectors is \textasciitilde{}1.0 kV higher than that with the standard ALICE mixture. A similar effect has been already obtained in the gas mixture of 45\% C$_{3}$H$_{2}$F$_{4}$ and 55\% O$_{2}$, as shown in figure \ref{fig:plot_misto}. The streamer probability in the gas mixture with 55\% CO$_{2}$ is \textasciitilde{}50\% at 90\% efficiency, as shown in figure \ref{fig:plot_anidridecarbonica}b. This value is lower than the streamer probability obtained in C$_{3}$H$_{2}$F$_{4}$-based gas mixtures with the addition of 55\% Ar, N$_{2}$ or O$_{2}$, while it is still far from the value in the standard ALICE mixture. Indeed, the streamer probability with 45\% C$_{3}$H$_{2}$F$_{4}$ and the remaining fraction of Ar, N$_{2}$ or O$_{2}$ is \textasciitilde{}80\% at 90\% efficiency, while it is less than 5\% in the standard ALICE mixture. In the mixture with 55\% CO$_{2}$, the average signal charge is \textasciitilde{}27 pC at the working point (11.2 kV), which is one order of magnitude greater than that of the standard ALICE mixture.

If the fraction of CO$_{2}$ is increased and that of C$_{3}$H$_{2}$F$_{4}$ is decreased, the working point turns out to be shifted towards lower voltages. Increasing the CO$_{2}$ fraction from 45\% to 55\% reduces the working point by \textasciitilde{}1 kV. Figure \ref{fig:plot_anidridecarbonica}b shows the streamer probability as a function of $HV$\textminus$HV_{\epsilon=0.9}$. The increase of C$_{3}$H$_{2}$F$_{4}$ fraction against CO$_{2}$ reduces streamers. In fact, going from 45\% to 55\% C$_{3}$H$_{2}$F$_{4}$, the streamer probability at 90\% efficiency decreases from \textasciitilde{}40\% to \textasciitilde{}20\%.

Taking into account the RPC working point, the streamer probability and the average charge of signals, CO$_{2}$ turns out to be the most promising solution in comparison to Ar, N$_{2}$ and O$_{2}$. However, the streamer probability is quite high in gas mixtures with only C$_{3}$H$_{2}$F$_{4}$ and CO$_{2}$, so results with the addition of \textit{i}-C$_{4}$H$_{10}$, which has a high photo-absorption cross section, and addition of SF$_{6}$, which is a strong electronegative gas, are presented in the following.

\subsection{C$_{3}$H$_{2}$F$_{4}$-based gas mixtures with addition of CO$_{2}$ and \textit{i}-C$_{4}$H$_{10}$}

The use of \textit{i}-C$_{4}$H$_{10}$ in the C$_{2}$H$_{2}$F$_{4}$-based gas mixtures for RPCs helps to reduce the photon-feedback effects, thanks to its high photo-absorption cross section \cite{sauli_ok}.

\begin{figure}[h]
    \centering
    \includegraphics[width=0.80\textwidth]{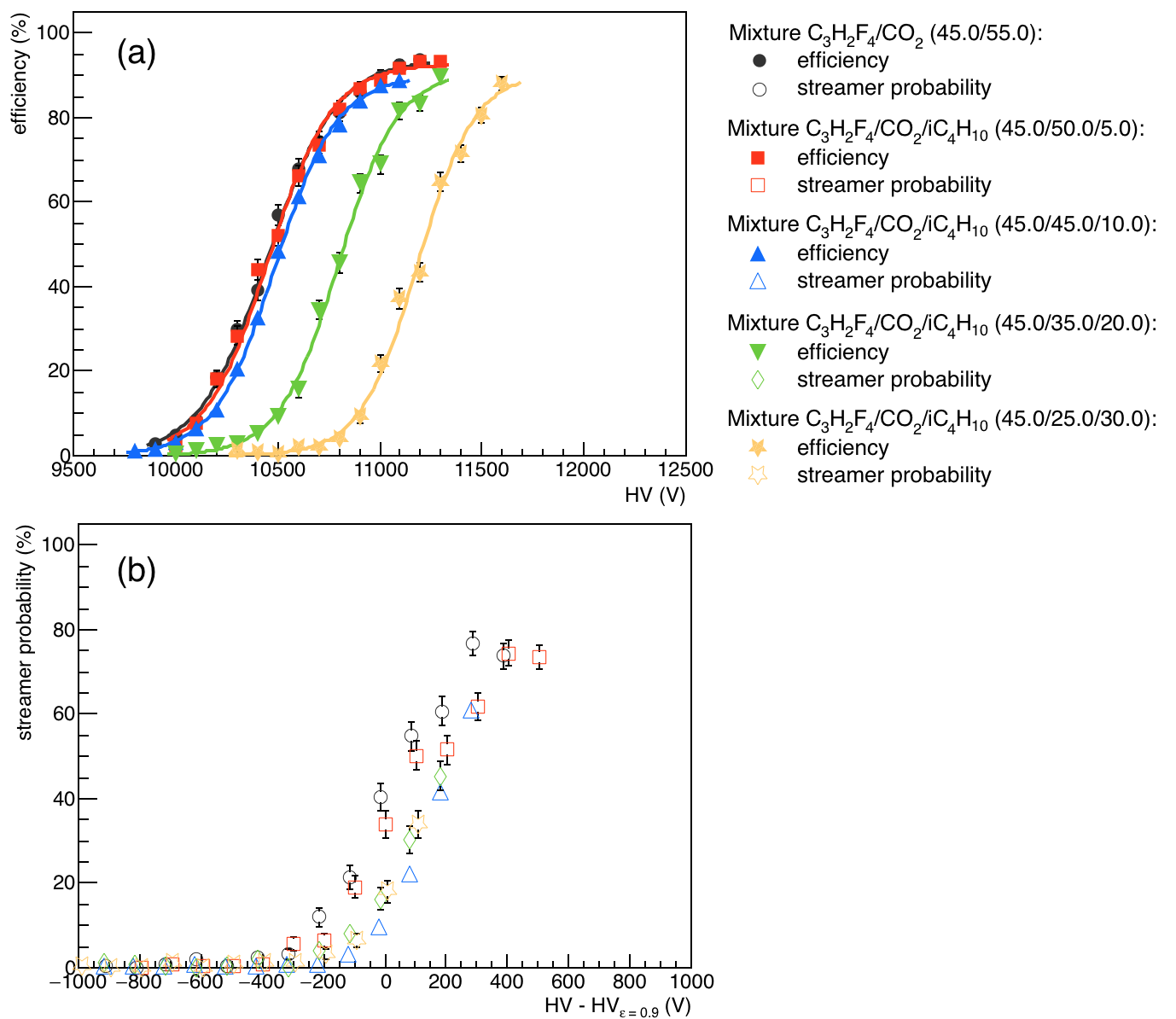}
    \caption{Efficiency (a) and streamer probability (b) of the RPC with 45\% C$_{3}$H$_{2}$F$_{4}$ and different ratios between CO$_{2}$ and \textit{i}-C$_{4}$H$_{10}$. Some statistical error bars are hidden by markers.}
    \label{fig:plot_isobutano}
\end{figure}

Figures \ref{fig:plot_isobutano}a and \ref{fig:plot_isobutano}b show the efficiency and streamer probability curves, respectively, at different ratios between \textit{i}-C$_{4}$H$_{10}$ and CO$_{2}$, while the concentration of C$_{3}$H$_{2}$F$_{4}$ is kept constant at 45\%. The working point increases by \textasciitilde{}600 V when going from 10\% to 20\% \textit{i}-C$_{4}$H$_{10}$, while no clear trend emerges when the \textit{i}-C$_{4}$H$_{10}$ fraction is varied between 0\% and 10\%. The streamer suppression is more efficient with \textit{i}-C$_{4}$H$_{10}$ concentrations higher than 10\%. Indeed, streamer probability at 90\% efficiency is \textasciitilde{}40\% without \textit{i}-C$_{4}$H$_{10}$ or with 5\% \textit{i}-C$_{4}$H$_{10}$, while it is lower than 20\% with more than 10\% \textit{i}-C$_{4}$H$_{10}$.

\subsection{C$_{3}$H$_{2}$F$_{4}$-based gas mixtures with addition of CO$_{2}$ and SF$_{6}$}

The effect of 1\% SF$_{6}$ has been tested in a gas mixture with only C$_{3}$H$_{2}$F$_{4}$ and CO$_{2}$. Thanks to its high attachment cross sections \cite{christophorou2001esafluoruro_ok}, the addition of SF$_{6}$ in C$_{2}$H$_{2}$F$_{4}$-based gas mixtures allows the suppression of streamers \cite{camarri1998streamer_ok}.

\begin{figure}[h]
    \centering
    \includegraphics[width=0.80\textwidth]{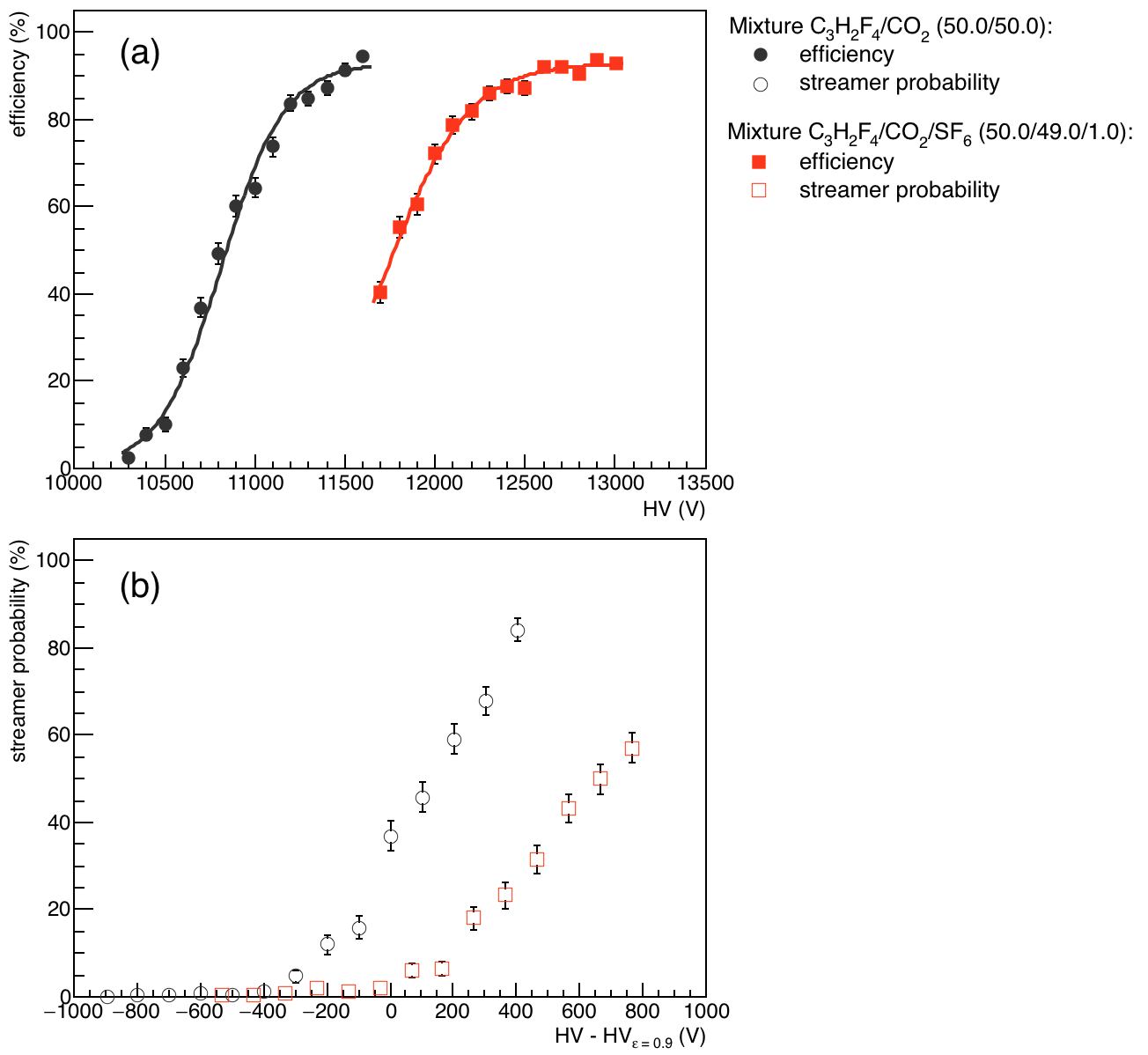}
    \caption{Efficiency (a) and streamer probability (b) of the RPC in C$_{3}$H$_{2}$F$_{4}$-based gas mixtures and CO$_{2}$ with and without the addition of 1\% SF$_{6}$. Some statistical error bars are hidden by markers.}
    \label{fig:plot_esafluoruro}
\end{figure}

Figure \ref{fig:plot_esafluoruro} shows how the efficiency curve and the streamer probability change if 1\% SF$_{6}$ is added to a gas mixture made of half C$_{3}$H$_{2}$F$_{4}$ and half CO$_{2}$. The shift of the working point is \textasciitilde{}1 kV towards higher voltages with the addition of 1\% SF$_{6}$. As expected, the streamers are greatly suppressed with SF$_{6}$. Indeed, the streamer probability at 90\% efficiency is \textasciitilde{}40\% without SF$_{6}$, whereas it is reduced down to \textasciitilde{}5\% with 1\% SF$_{6}$.

The addition of 1\% SF$_{6}$ reduces also the average charge of signals. At the working point, the average charge is \textasciitilde{}33 pC in the gas mixture of 50\% C$_{3}$H$_{2}$F$_{4}$ and 50\% CO$_{2}$, while it becomes only \textasciitilde{}4 pC in the gas mixture with 1\% SF$_{6}$.

\section{Conclusions}\label{sec:conclusions}
The basic performance of C$_{3}$H$_{2}$F$_{4}$-based gas mixtures for RPCs have been studied. The usage of 100\% C$_{3}$H$_{2}$F$_{4}$ in RPCs is not advisable because it leads to a too high operating voltage, as discussed in section \ref{sec:metodologia}.

The addition of Ar, N$_{2}$, O$_{2}$ and CO$_{2}$ in C$_{3}$H$_{2}$F$_{4}$-based gas mixtures reduces the operating voltage of the detector. However, the use of Ar, N$_{2}$ and O$_{2}$ in C$_{3}$H$_{2}$F$_{4}$-based gas mixtures implies high streamer probability and signal charge at the operating voltage. In addition, current instability is observed with the use of O$_{2}$ in the gas mixture.

CO$_{2}$ turns out to be the most promising solution. Indeed, values of streamer probability and signal charge are lower than those obtained in C$_{3}$H$_{2}$F$_{4}$-based gas mixtures with addition of Ar, N$_{2}$ and O$_{2}$, as shown in section \ref{sec:characterization_gases}. No current instability has been observed in any of the C$_{3}$H$_{2}$F$_{4}$-based gas mixtures with CO$_{2}$, tested so far. Measurements with gas mixtures made of C$_{3}$H$_{2}$F$_{4}$ and CO$_{2}$ show that the addition of both \textit{i}-C$_{4}$H$_{10}$, as quencher, and SF$_{6}$, as strong electronegative gas, improves the RPC performance.

The interplay of C$_{3}$H$_{2}$F$_{4}$, CO$_{2}$, \textit{i}-C$_{4}$H$_{10}$ and SF$_{6}$ has been already investigated for single-gap RPCs in avalanche mode \cite{bianchi2019characterization_ok}. The most promising gas mixtures, which have been found so far \cite{bianchi2019characterization_ok}, consist of $50\%$ CO\textsubscript{2}, $39.7\%$ C$_{3}$H$_{2}$F$_{4}$, $10\%$ \textit{i-}C\textsubscript{4}H\textsubscript{10}, $0.3\%$ SF\textsubscript{6} and $50\%$ CO\textsubscript{2}, $39\%$ C$_{3}$H$_{2}$F$_{4}$, $10\%$ \textit{i-}C\textsubscript{4}H\textsubscript{10}, $1\%$ SF\textsubscript{6}. Their GWPs are respectively 72 and 232, while the GWP of gas mixtures for RPCs in muon detection systems of the LHC experiments is \textasciitilde{}1400.

Measurements of the RPC performance under irradiation are ongoing at the CERN Gamma Irradiation Facility \cite{gif_ok} in order to assess the rate capability and ageing properties resulting from the C$_{3}$H$_{2}$F$_{4}$-based gas mixtures.



\end{document}